\documentclass[twocolumn,english,prl,superscriptaddress,showpacs,amssymb]{revtex4}
\usepackage[T1]{fontenc}
\usepackage{amsmath}
\usepackage{amssymb}
\usepackage{graphicx}

\makeatletter
\@ifundefined{textcolor}{}
{%
 \definecolor{BLACK}{gray}{0}
 \definecolor{WHITE}{gray}{1}
 \definecolor{RED}{rgb}{1,0,0}
 \definecolor{GREEN}{rgb}{0,1,0}
 \definecolor{BLUE}{rgb}{0,0,1}
 \definecolor{CYAN}{cmyk}{1,0,0,0}
 \definecolor{MAGENTA}{cmyk}{0,1,0,0}
 \definecolor{YELLOW}{cmyk}{0,0,1,0}
 }

\usepackage{bm}\@ifundefined{definecolor}{\usepackage{color}}{}

\newcommand{\bh}[1]{#1}

\usepackage{babel}

\makeatother

\usepackage{babel}
\begin{document}

\title{Widely tunable, non-degenerate three-wave mixing microwave device
operating near the quantum limit}

\author{N. Roch}
\affiliation{Laboratoire Pierre Aigrain, Ecole Normale Sup\'erieure, CNRS (UMR 8551),
Universit\'e P. et M. Curie, Universit\'e D. Diderot
24, rue Lhomond, 75231 Paris Cedex 05, France
}
\author{E. Flurin}
\affiliation{Laboratoire Pierre Aigrain, Ecole Normale Sup\'erieure, CNRS (UMR 8551),
Universit\'e P. et M. Curie, Universit\'e D. Diderot
24, rue Lhomond, 75231 Paris Cedex 05, France
}
\author{F. Nguyen}\thanks{Present address: National Institute of Standards and Technology, 325 Broadway, Boulder CO 80305}

\affiliation{Laboratoire Pierre Aigrain, Ecole Normale Sup\'erieure, CNRS (UMR 8551),
Universit\'e P. et M. Curie, Universit\'e D. Diderot
24, rue Lhomond, 75231 Paris Cedex 05, France
}

\author{P. Morfin}
\affiliation{Laboratoire Pierre Aigrain, Ecole Normale Sup\'erieure, CNRS (UMR 8551),
Universit\'e P. et M. Curie, Universit\'e D. Diderot
24, rue Lhomond, 75231 Paris Cedex 05, France
}

\author{P. Campagne-Ibarcq}
\affiliation{Laboratoire Pierre Aigrain, Ecole Normale Sup\'erieure, CNRS (UMR 8551),
Universit\'e P. et M. Curie, Universit\'e D. Diderot
24, rue Lhomond, 75231 Paris Cedex 05, France
}
\author{M. H. Devoret}

\affiliation{Coll\`ege de France, 11 Place Marcelin Berthelot, F-75231 Paris Cedex 05, France
}
\affiliation{Laboratoire Pierre Aigrain, Ecole Normale Sup\'erieure, CNRS (UMR 8551),
Universit\'e P. et M. Curie, Universit\'e D. Diderot
24, rue Lhomond, 75231 Paris Cedex 05, France
}
\affiliation{Department of Applied Physics, Yale University,
PO Box 208284, New Haven, CT 06520-8284
}

\author{B. Huard}\email[corresponding author: ]{benjamin.huard@ens.fr} 
\affiliation{Laboratoire Pierre Aigrain, Ecole Normale Sup\'erieure, CNRS (UMR 8551),
Universit\'e P. et M. Curie, Universit\'e D. Diderot
24, rue Lhomond, 75231 Paris Cedex 05, France
}

\date{\today}
\begin{abstract}
We present the first experimental realization of a widely frequency
tunable, non-degenerate three-wave mixing device for quantum signals
at GHz frequency. It is based on a new superconducting building-block
consisting of a ring of four Josephson junctions shunted by a cross
of four linear inductances. The phase configuration of the ring remains
unique over a wide range of magnetic fluxes threading the loop. It
is thus possible to vary the inductance of the ring with flux while
retaining a strong, dissipation-free, and noiseless non-linearity.
The device has been operated in amplifier mode and its noise performance
has been evaluated by using the noise spectrum emitted by a voltage
biased tunnel junction at finite frequency as a test signal. The unprecedented
accuracy with which the crossover between zero-point-fluctuations
and shot noise has been measured provides an upper-bound for the noise
and dissipation intrinsic to the device. 
\end{abstract}
\maketitle
Three-wave mixing devices, i.e. non-linear circuits converting power
among three microwave signals, are key elements of analog information
processing in the microwave domain\cite{Pozar}. However, they are
based on dissipative components such as semiconductor diodes, or SIS
tunnel junctions biased near the superconducting gap\cite{Tinkham}.
The loss of signal limits their operation and also introduces noise
above the minimum required by quantum mechanics\cite{CAVES:1982p1120,Clerk:2010p7260}. A non-degenerate mixing device with noise
close to that minimum level  was demonstrated recently~\cite{Bergeal:2010p985,Bergeal:2010p984}. However, the hysteresis  preventing  flux tunability for this 4-junction circuit severely limited possible applications to analog quantum signal processing. In this Letter,
we show that by adding four inductances to the 4-junction loop, we can fully suppress the hysteresis and reach a 500MHz  frequency tunability while operating close to the quantum limit. Our improvement of the device tunability by an order of magnitude is obtained  without jeopardizing other advantages of non-degenerate 3-wave mixing.

\begin{figure}
\includegraphics[scale=0.4]{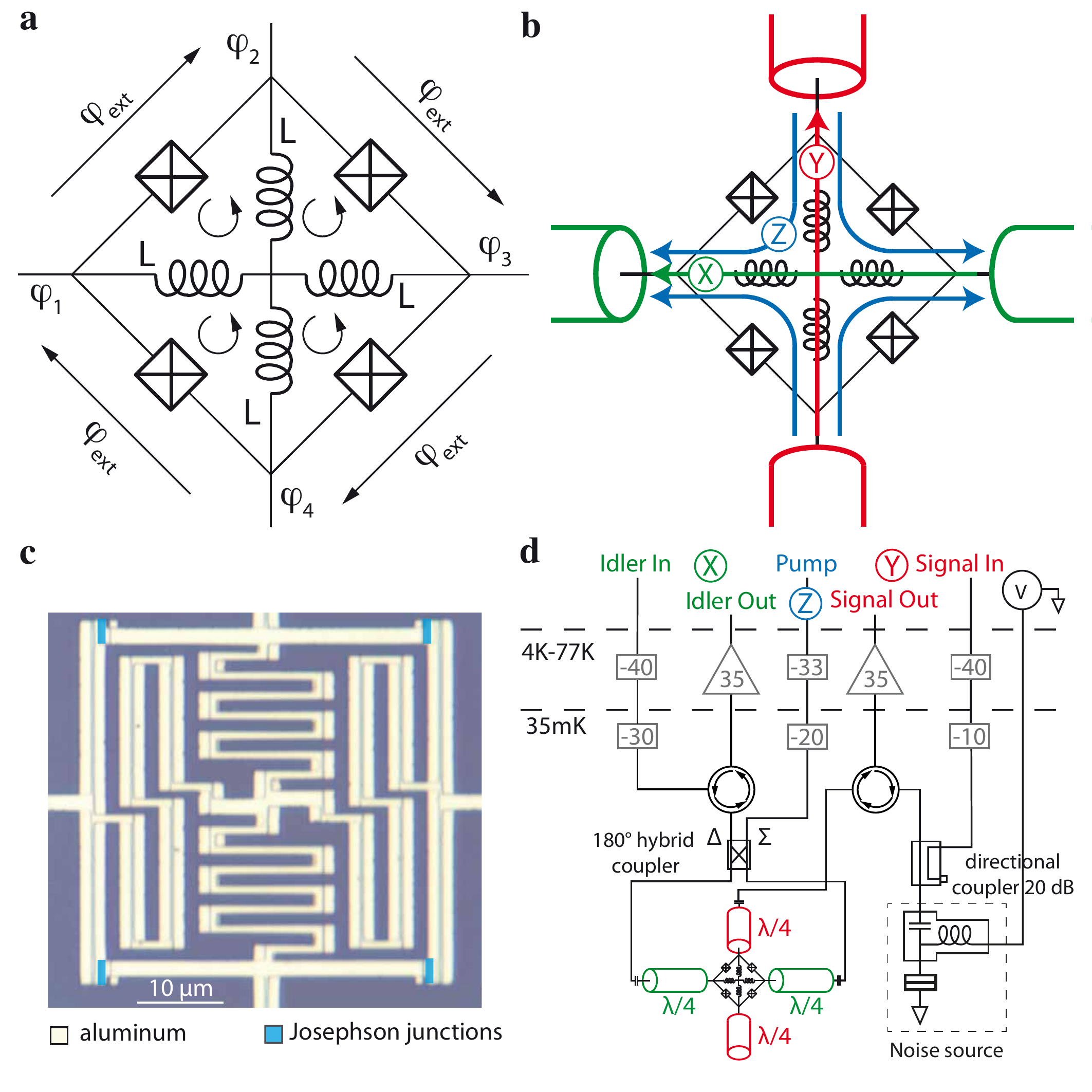} \caption{\textbf{a}. Device schematic: four linear inductances $L$ cross-link
a ring of four Josephson junctions. Each sub-loop is biased by a magnetic
flux $\varphi_{ext}\varphi_{0}$. For $L<L^0_J/4$,
the current through the inductances is zero and the external flux
phase biases the junctions to $\varphi_{ext}$. \textbf{b}. The device
is embedded at the intersection of four transmission lines and couples
to spatial modes $X$, $Y$ and $Z$ represented as arrows. \textbf{c}.
Optical microscope image of the ring modulator. The meanders in the center of the ring implement the four linear inductances from \textbf{a}. The stripes on the meanders are due to the fabrication process based on shadow evaporation. \textbf{d}. Simplified
schematic of the setup used to characterize 3-wave mixing operation.
The idler resonator ($X$) is excited through a $180^{\circ}$ hybrid
coupler while the signal resonator ($Y$) is single-ended. The noise
emitted by the voltage biased tunnel junction in its normal state
is amplified through the signal port.}

\label{figure1} 
\end{figure}

An ideal non-degenerate three-wave mixing device in the microwave
domain absorbs three signals at frequencies such that $\omega_{X}+\omega_{Y}=\omega_{Z}$
with complex amplitudes $A_{X}^{in}$, $A_{Y}^{in}$ and $A_{Z}^{in}$,
respectively, and reemits signals at the same frequencies with amplitudes
$A_{X}^{out}$, $A_{Y}^{out}$ and $A_{Z}^{out}$ such that $\left\vert A_{X}^{out}\right\vert ^{2}+\left\vert A_{Y}^{out}\right\vert ^{2}+\left\vert A_{Z}^{out}\right\vert ^{2}=\left\vert A_{X}^{in}\right\vert ^{2}+\left\vert A_{Y}^{in}\right\vert ^{2}+\left\vert A_{Z}^{in}\right\vert ^{2}$,
that is without internal dissipation. The device can operate in two
power amplification modes: i) the photon gain mode, for which $\left\vert A_{Z}^{in}\right\vert ^{2}\gg\left|A_{X}^{in}\right|^{2}$,
$\left|A_{Y}^{in}\right|^{2}$ is the pump power providing the extra
photon numbers in the re-emitted signals at frequencies $\omega_{X}$
and $\omega_{Y}$, and ii) the pure up-conversion mode for which $\left\vert A_{Y}^{in}\right\vert ^{2}\gg A_{X}^{in}$,
$A_{Z}^{in}$ is the pump power providing the energy difference between
photons at $\omega_{Z}$ and photons at $\omega_{X}$. The Josephson
Parametric Converter (JPC)\cite{Bergeal:2010p984}, consisting of
a ring of four Josephson junctions, can perform both functions. However,
its operation has little tunability since the flux $\Phi_{ext}$ applied
through the ring has to be adjusted in the close vicinity of the value
$\Phi_{0}/2$, where $\Phi_{0}=h/2e$ is the flux quantum. In the
present work, we consider a more general 3-wave mixing device in which
4 linear inductances are cross-linking the ring-modulator like the
spokes of a wheel (see Fig.\ref{figure1}\textbf{a}). The hamiltonian of the
ring is

\begin{equation}
\begin{array}{rcl}
H_{{}} & = & -\frac{1}{2}E_{J}\sin{(\varphi_{ext})}\varphi_{X}\varphi_{Y}\varphi_{Z}\\
 &  & +\frac{1}{2}\left(E_{L}/2+E_{J}\cos{\varphi_{ext}}\right)\left({\varphi_{X}}^{2}+{\varphi_{Y}}^{2}\right)\\
 &  & +\frac{1}{2}\left(E_{L}/4+E_{J}\cos{\varphi_{ext}}\right){\varphi_{Z}}^{2}+O({\varphi_{X,Y,Z}}^{4})
\end{array}\label{hamiltonian}
\end{equation}

where the three spatial mode amplitudes $\varphi_{X}=\varphi_{1}-\varphi_{3}$,
$\varphi_{Y}=\varphi_{2}-\varphi_{4}$ and $\varphi_{Z}=\varphi_{1}+\varphi_{3}-\varphi_{2}-\varphi_{4}$
are gauge-invariant, orthogonal linear combinations of the superconducting
phases of the four nodes of the Josephson junction ring (Fig. \ref{figure1}\textbf{b}).

We will see below how these standing wave modes can be excited by
the propagating mode amplitudes $A_{X}^{in}$, $A_{Y}^{in}$ and $A_{Z}^{in}$
and emit the amplitudes $A_{X}^{out}$, $A_{Y}^{out}$ and $A_{Z}^{out}$.
In the hamiltonian (\ref{hamiltonian}), $E_{L}={\varphi_{0}}^{2}/L$
is the energy associated with each of the inductances $L$, and  $E_{J}=\varphi_0^2/L^0_J$ is the Josephson
energy of each tunnel junction. We also
define the reduced flux quantum $\varphi_{0}=\hbar/2e$ and the dimensionless
flux $\varphi_{ext}=\Phi_{ext}/4\varphi_{0}$ threading each of the
nominally identical 4 loops of the device. The first term of the hamiltonian
is a pure 3-wave mixing term, while the two others are quadratic terms
determining the effective inductance of modes $X$, $Y$ and $Z$
: $L_{X,Y,Z}^{-1}=\varphi_{0}^{-2}\partial^{2}H/\partial\varphi_{X,Y,Z}^{2}$.
The value ${\varphi_{ext}=}\pi/2$ maximizes the strength of the mixing
term. Provided that $E_{L}/2>E_{J}$, modes $X$ and $Y$ can be tuned
by varying ${\varphi_{ext}}$ while retaining their stability: $L_{X,Y}^{-1}>0$
on the whole range of variation. However, there is a range of fluxes
for which $L_{Z}^{-1}<0$ where the device departs from $\left\langle \varphi_{Z}\right\rangle =0$
so that the expansion (\ref{hamiltonian}) is inappropriate. If the
inductances are lowered even more such that $E_{L}/4>E_{J}$, then
all three modes of the device are stable for every value of $\varphi_{ext}$,
but at the expense of significant dilution of the non-linear term.
In contrast, as $E_{L}$ is lowered below $2E_{J}$, dilution of non-linearity
is minimized, but at the expense of the stability of the three modes.
This is why the JPC, for which $E_{L}=0$, can operate only within
a small range of values of ${\varphi_{ext}}$ forbidding any tunability
of the device.

We have tested this new, tunable, mixing element design, by inserting
the ring into a resonant structure consisting of two $\lambda/2$
transmission line resonators coupled to the $X$ and $Y$ modes (Fig.
\ref{figure1}\textbf{d}) as in Ref.~\cite{Abdo:2011p1063}. The $Z$ mode is non-resonant and excited through resonator X using a hybrid coupler (Fig. \ref{figure1}\textbf{d}).
By varying the externally applied flux, it is possible to adjust the
$X$ and $Y$ resonator frequencies given by 
\begin{equation}
\omega_{X,Y}=\omega_{X,Y}^{0}\frac{\pi^{2}L_{X,Y}^{\lambda/2}/2}{\pi^{2}L_{X,Y}^{\lambda/2}/2+L_{X,Y}(\varphi_{ext})},\label{freqvary}
\end{equation}
 where $\omega_{X,Y}^{0}$ is the resonance frequency of the bare
$\lambda/2$ resonator without a ring, $L_{X,Y}^{\lambda/2}=2Z_{0}/(\pi\omega_{X,Y}^{0})$
its lumped-element equivalent inductance\cite{Pozar} and $Z_{0}$
its characteristic impedance. As long as $E_{L}/4+E_{J}\cos{\varphi_{ext}}>0$,
the ring inductance $L_{X,Y}$ is given by

\begin{equation}
L_{X,Y}(\varphi_{ext})=\varphi_{0}^{2}\left(\frac{E_{L}}{2}+E_{J}\cos\varphi_{ext}\right)^{-1}.\label{eq:inductance}
\end{equation}

\begin{figure}
\includegraphics[scale=0.7]{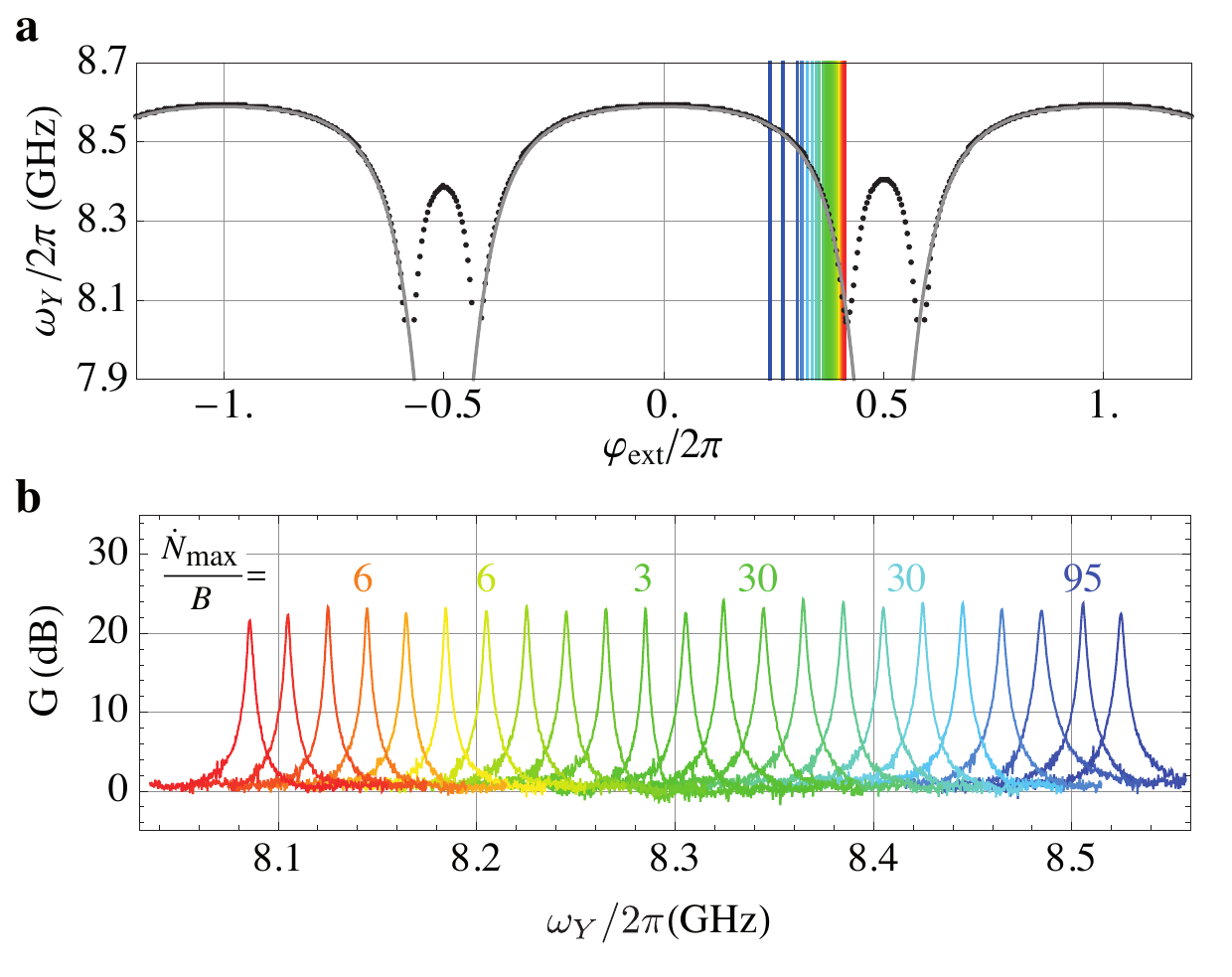} \caption{\textbf{a}. Dots: Measured resonance frequency $\omega_{Y}$ of the signal cavity
as a function of flux applied to the ring modulator without pump. Solid line: fit of $\omega_{Y}$ using Eq. (\ref{freqvary}) with $\omega^0_Y/2\pi=8.82~\mathrm{GHz}$, $L=49~\mathrm{pH}$, $E_J=\varphi_0\times 1.9~\mu\mathrm{A}$ and including the known stray inductance around the loop $4L_S=200~\mathrm{pH}$ (see ref.\cite{note1}).
\textbf{b}. Reflection gain measured
on the signal port as a function of frequency for various values of
the flux indicated by the color lines in \textbf{a}. Pump parameters
are optimized for each curve. The numbers on top represent the $1~\text{dB}$
compression point (maximum input power) expressed in input photon
rate per dynamical bandwidth for six different working frequencies
coded by color.}

\label{figure2} 
\end{figure}

The device presented in Fig. \ref{figure1}\textbf{c} is realized in a single
e-beam lithography step. The critical current of the Al/Al$_{2}$O$_{3}$/Al
Josephson junctions was designed to be in the $\mathrm{\mu A}$ range.
The wide geometric linear inductances cross-linking the ring are approximately
given by $\mu_{0}l$ where $l=100~\mu\mathrm{m}$ is the length
of each of the four meanders. According to theory, they should present
negligible kinetic inductance\cite{Annunziata:2010p7376}. The value
of the ratio $E_{L}/E_{J}=3\pm 2$ should favor the stability
of the X and Y modes.

The device was operated in the photon gain mode. The phase and amplitude of the waves $A_{X}^{out}$ and $A_{Y}^{out}$,
relative to those of $A_{X}^{in}$ and $A_{Y}^{in}$ are measured
with a vector network analyzer, for a whole set of pump tones $A_{Z}^{in}$.
Turning off the pump tone first, we obtained the resonance frequency
of both resonators as a function of flux (see Fig. \ref{figure2}\textbf{a})
as well as their half-maximum bandwidths $B_{X}=39~\mathrm{MHz}$
and $B_{Y}=29~\mathrm{MHz}$. Unlike in the JPC, no hysteresis was
found in the dependence of the resonance frequency on applied flux,
confirming the stability of our device. However, two regimes must
be distinguished in the data: that of the wide arches obeying (\ref{freqvary})
with a ring inductance given by (\ref{eq:inductance}) and that of
the narrow arches for which $E_{L}/4+E_{J}\cos\varphi_{ext}<0$ and
where the ring inductance depends precisely on the non-zero value
of $\left\langle \varphi_{Z}\right\rangle $ emerging from the broken
symmetry along the $Z$ mode. It is interesting to note that the two
possible opposite values for $\left\langle \varphi_{Z}\right\rangle $
in this regime give exactly the same resonance frequency. Besides, the fit of Fig.~\ref{figure2}\textbf{a} does not take into account the perturbative effect of the parasitic inductances in series with the junctions. Using the full
hamiltonian and these stray inductances, a complete
agreement with the data can be obtained over the full flux variation range~\cite{note1}.

\begin{figure}
\includegraphics[scale=0.35]{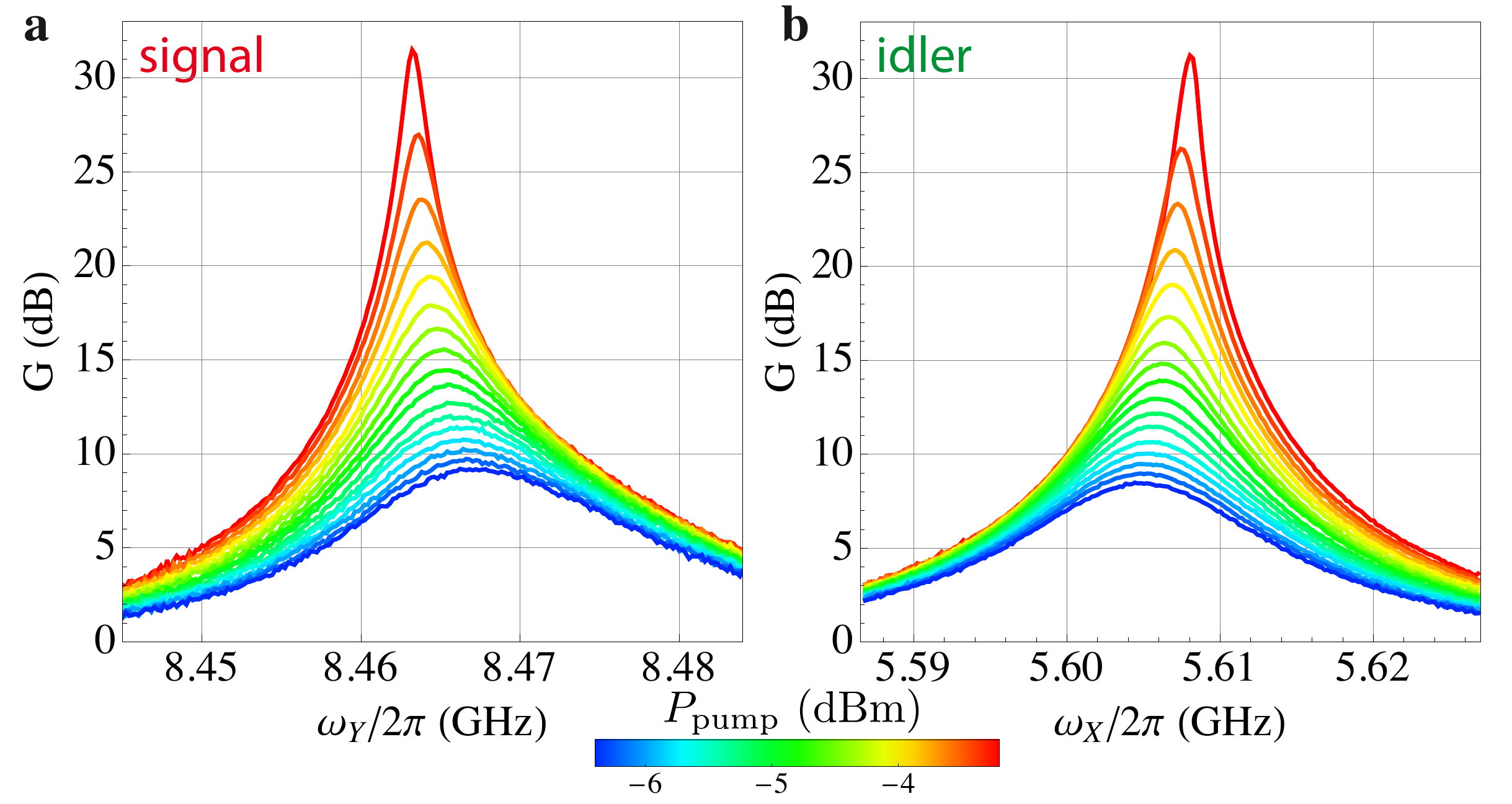} \caption{Reflection gain of the phase-preserving Josephson amplifier observed
on the signal Y (\textbf{a}) and the idler X (\textbf{b}) modes. The
color bar indicates the pump power referred to the output of the generator.
The pump frequency is $\omega_{Z}/2\pi=14.071~\mathrm{GHz}$ and the
flux is set to $\varphi_{ext}/2\pi\approx0.3125$. \label{figure3}}
\end{figure}

The power gain $G$ of the device is defined as the ratio of the reflected
power with pump on and off. The dependence of the gain on the pump
power is shown on Fig.~\ref{figure3}. Note in particular that a
dynamical bandwidth $B=3.2\ \mathrm{MHz}$ is obtained for a gain
of $20\ \mathrm{dB}$. We checked that the parametric amplifier relation
$\sqrt{G}\times B(G)=2\left(B_{X}^{-1}+B_{Y}^{-1}\right)^{-1}$ holds
to less than a MHz of deviation for any pump power yielding a gain
greater than 5~dB, for both signal and idler waves, as theory predicts \cite{Bergeal:2010p985}.

As illustrated on Fig.~\ref{figure2}\textbf{b, }the amplifier center
frequency can be flux-tuned over $400~\mathrm{MHz}$ which represents
a range two orders of magnitude greater than the bandwidth at 20dB.
Indeed, for each center frequency, we can find a reproducible set
of applied flux, pump power and pump frequency yielding a gain higher
than $20\ \mathrm{dB}$ and a dynamical bandwidth of $B=3~\mathrm{MHz}$
(Fig~\ref{figure2}\textbf{b}). No amplification was found in the
domain of the narrow arches. While this observation cannot be explained
directly by the expansion (\ref{hamiltonian}), it is consistent with
the full hamiltonian that predicts the non-linear term to be significantly
spoiled by spurious terms when $\left\langle \varphi_{Z}\right\rangle \neq0$.
The key point of our experiment is that we can still benefit, outside
the range of the narrow arches, from a confortable tunable 3-wave
non-linearity. The tunability of this non-degenerate amplifier can therefore compete with the
state of the art degenerate Josephson amplifiers\cite{CastellanosBeltran:2008p705,Hatridge:2011p1129,Yamamoto:2008p708,Vijay:2011p1051,Eichler,Wilson:2010p1081}
with the added benefits of pump-signal separation.

We now turn to dynamical range measurements which further characterize
the non-linear operation of our device. For these measurements, we
first calibrated the attenuation of the line named \textquotedbl{}Signal
In\textquotedbl{} (Fig.~\ref{figure1}\textbf{d}) with an accuracy
of 3 dB\cite{note2}. We then measured the so-called $1~\text{dB}$
compression point of the amplifier mode of our device, which is the
input power for which the gain is reduced by $1~\text{dB}$. As presented
on Fig.~\ref{figure2}\textbf{b}, this maximal power ranges between
$-133~\text{dBm}$ and $-118~\text{dBm,}$ corresponding to 3 and
95 photons per inverse dynamical bandwidth. The reduction in maximal
allowed power occurs at lower frequencies where we have also observed
that the pump power needed for a given gain is $\sim30~\text{dB}$
lower than at higher frequencies. We believe that it could be explained
by the pump frequency becoming, at lower signal frequencies, resonant
with a mode of the crossed resonators. The device would hence depart
from the stiff pump condition needed for parametric amplification
with maximal dynamic range.

\begin{figure}
\includegraphics[scale=0.37]{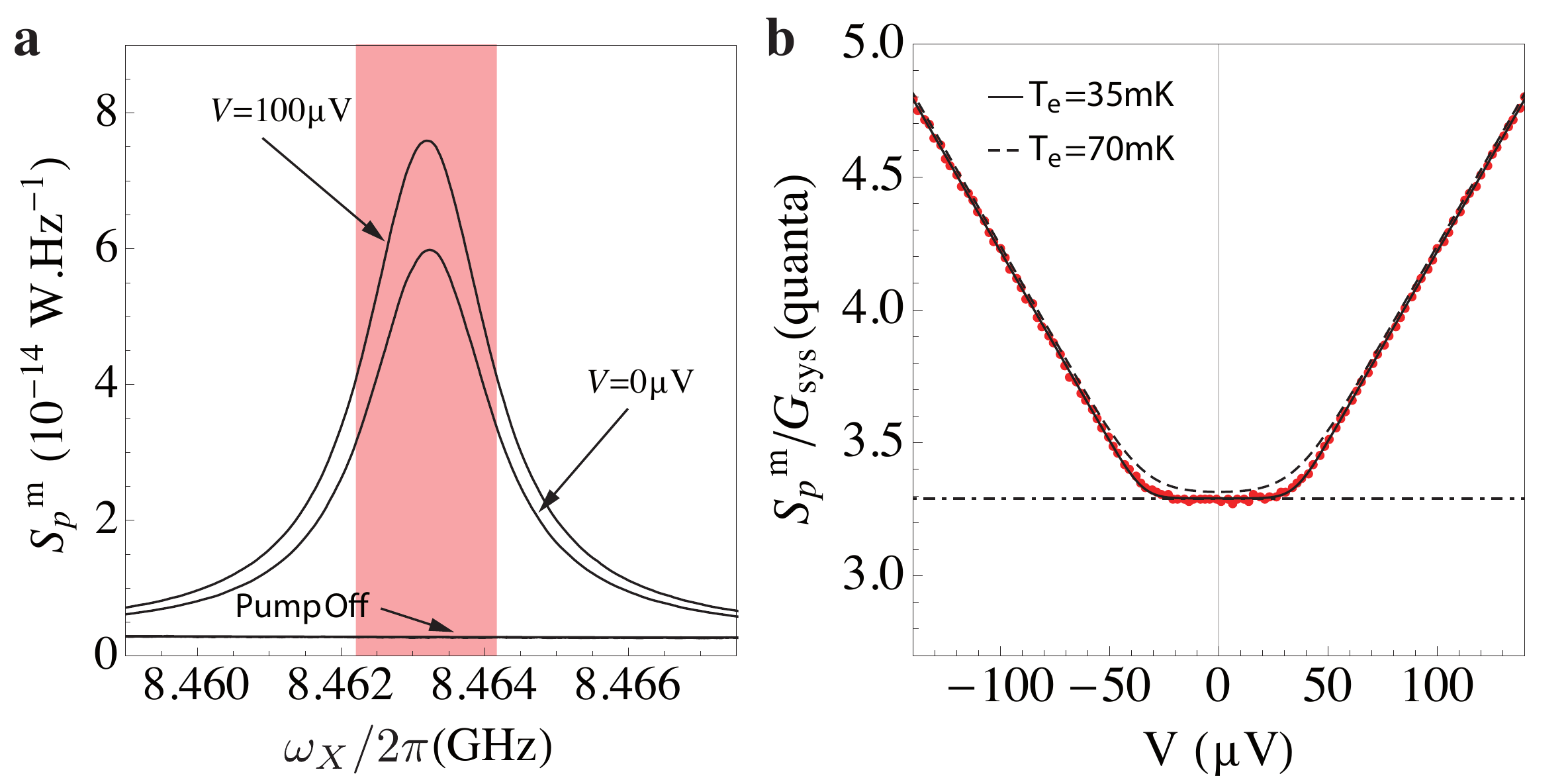} \caption{\textbf{a}. Power spectral density as a function of frequency measured
at the output of \textquotedbl{}Signal Out\textquotedbl{} for three
settings: pump \emph{off} and $V=0$, pump \emph{on} and $V=0$ or
$V=100~\mu$V. The colored area represents the averaging range used
in the right panel. \textbf{b}. Average power spectral density over
a $2\ \mathrm{MHz}$ bandwidth around the center frequency of the
amplifier as a function of bias voltage $V$. The solid line shows
what is expected using Eq.~(\protect\ref{totalnoise}) and fitting
an overall gain $G_{sys}=94.6~\mathrm{dB}$ and an extra noise $N_{add}$
of 2.8 quanta coming both from the unavoidable quantum noise of the
idler port (0.5 quanta) and the unwanted losses between the tunnel
junction and the amplifier (2.3 quanta). The gain $G_{sys}$ allows
us to express this power spectral density in units of photon number
or quantum.}

\label{figure4} 
\end{figure}

In a last series of experiments, the noise of our device was assessed
by using the noise emitted by a voltage biased NIN tunnel junction
as input signal. This noise, which is well-understood and therefore
of predictable amplitude, plays the role of an \emph{in situ} calibrated
signal. At small electronic temperatures ($k_{B}T_{e}\ll\hbar\omega_{S}$),
the noise from a tunnel junction presents two regimes as a function
of voltage. For $eV<\hbar\omega_{S}$, zero-point fluctuations across
the junction dominate with a power spectral density ${S_{p}}(\omega_{S})=\frac{\hbar\omega_{S}}{2}$,
while for $eV>\hbar\omega_{S}$, electrons in the junction produce
non-equilibrium shot noise and ${S_{p}}(\omega_{S})=\frac{eV}{2}$.
The electronic temperature $T_{e}$ in the electrodes of the junction
sets the sharpness of the crossover between these two regimes \cite{Schoelkopf:1997p1121,Blanter:2000p1050}
as $S_{p}=S_{p}^{+}+S_{p}^{-}$ with

\begin{equation}
S_{p}^{\pm}(\omega)=\frac{1}{4}(eV\pm\hbar\omega)\coth{\frac{eV\pm\hbar\omega}{2k_{B}T_{e}}}.\label{PSDjunction}
\end{equation}

Our experiment was performed using an aluminum junction kept in its
normal state by permanent magnets close-by. We measured \emph{in situ}
a normal resistance of $43.9\ \mathrm{\Omega}$ (measurement lines
not shown on Fig.~\ref{figure1}\textbf{b}). The output spectral
density was recorded with a spectrum analyzer and averaged over a
$2\ \mathrm{MHz}$ bandwidth around the center frequency of the amplifier
(see Fig.~\ref{figure4}\textbf{a}). Its dependence with bias voltage was
obtained (Fig.~\ref{figure4}\textbf{b}) for an amplifier gain of $23\ \mathrm{dB}$
with the same settings as in Fig.~\ref{figure3}. The measured power
spectral density is remarkably well described by an expression of
the form: 
\begin{equation}
{S_{p}}^{m}(\omega_{S})=G_{sys}(S_{p}+N_{add}\ \hbar\omega_{S}).\label{totalnoise}
\end{equation}
 In the shot noise regime, it is possible to calibrate the system
gain $G_{sys}=\textrm{ d}{S_{p}}^{m}/\textrm{d}(eV/2)=94.6~\mathrm{dB}$
from the NIN tunnel element to the spectrum analyzer including a possible attenuation 
from the element to the input port of the amplifier. Without any additional
calibration, we extracted the apparent system added noise $N_{add}=2.8$ at
the plateau (Fig. \ref{figure4}\textbf{b}). This number of quanta can be thought
of as the standard half quantum attributable to the unavoidable quantum
noise of the load at the idler port, and 2.3 quanta left which can be seen as an upper bound on the
extra noise generated inside the device. On the other hand, an electronic
temperature $T_{e}$ equal to the refrigerator mixing chamber temperature of $35~\mathrm{mK}$
describes perfectly the crossover. It is worth emphasizing that the
noise power of the total measurement setup is presented in Fig.~\ref{figure4}
without any background subtraction and is therefore the full \emph{absolute}
\ system noise. In fact, there is a finite attenuation between the
junction and the amplifier leading to an underestimation of the gain counted from the input of amplifier  and hence to the actual noise added by the device. 
Besides the unwanted insertion loss inherent to our type of low temperature
measurement setup, the complex impedance of the junction itself is imperfectly
matched\cite{Spietz:2010p1064}. Given the size of the junction ($\simeq10\ \mathrm{{\mu m^{2}}}$)
and previous experiments on similar junctions, we estimated its capacitance
to be in the $0.7\mathrm{{pF}-1\ {pF}}$ range. Using the resistance
of the junction and the characteristic impedance of the amplifier,
we calculated that the loss of signal due to the RC filtering of the
junction noise leads to an apparent added noise between $1.3$ and
$2.1$ photons. Our measurement thus improves the $N_{add}$ found
by Bergeal \emph{et al.} in that the measurement frequency and bandwidth
are substantially higher\cite{Bergeal:2010p985}. It is straightforward
to compare the noise measurement with and without our device. Turning
off the pump tone, the same noise measurement using only a state-of-the-art
HEMT amplifier at $4~\mathrm{K}$\cite{Weinreb:1988p1126} yielded
an apparent added noise 20 times larger than with the pump on. This
translates into an acquisition time 400 times longer, keeping the
same bandwidth.

In conclusion, we have shown that it is possible to realize with Josephson
tunnel junctions a widely tunable, dissipation-less, non-degenerate
3-wave mixing element which processes microwave signals, adding a
level of noise not significantly greater than the level of unavoidable
quantum noise. Such an element could be useful in a certain number
of analog quantum signal processing applications, like the feedback
control of the state of a quantum bit\cite{Korotkov:2005p3727}.

Acknowledgments: Discussions with the Quantronics group at CEA Saclay,
as well as with F. Mallet, F. Schackert and T. Kontos have been greatly
useful. We gratefully acknowledge O. Andrieu and J.C. Dumont for technical
support. The devices have been made within the consortium Salle Blanche
Paris Centre. We thank D. Mailly for helping us dicing wafers at LPN-Marcoussis
and the Quantronics group for metal evaporations. This work was supported
by the EMERGENCES program Contract of Ville de Paris and by the ANR
contract ULAMSIG.

 \bibliographystyle{apsrev}

\begin{thebibliography}{18}
\bibitem[{not()}]{note2}
\bibinfo{note}{To calibrate \bh{the attenuation of} the "Signal In" line, we
  first measure the amplification of the "Signal Out" line \bh{using the tunnel
  junction spectrum. We can then easily deduce the attenuation at the input by
  measuring the transmission from "Signal In" to "Signal Out" lines over a wide
  range of powers. A 3~dB uncertainty remains due to the imperfectly known
  impedance of the tunnel junction.}}
\end{thebibliography}

\end{document}